\documentclass[
 reprint,
 superscriptaddress,
 nofootinbib,
 floatfix,
 amsmath,amssymb,
 aps,
 prl,
]{revtex4-2}
  
\usepackage{mathtools}
\usepackage[
    range-phrase=--,
    range-units=single,
    retain-explicit-plus=true,
    retain-unity-mantissa=false,
]{siunitx}
\usepackage{graphicx}
\usepackage{booktabs}
\usepackage[hidelinks]{hyperref}
\usepackage[nameinlink]{cleveref}

\AtBeginDocument{%
  \crefname{figure}{Fig.}{Figs.}
  \Crefname{figure}{Fig.}{Figs.}
  \crefname{equation}{Eq.}{Eqs.}
  \Crefname{equation}{Eq.}{Eqs.}
  \crefname{table}{Table}{Tables}
  \Crefname{table}{Table}{Tables}
  \crefname{section}{Sec.}{Secs.}
  \Crefname{section}{Sec.}{Secs.}
}
\usepackage[dvipsnames]{xcolor}

\newcommand{\Sh}{\ensuremath{S_h}}
\newcommand{\QCRB}{\ensuremath{\mathrm{QCRB}}}
\newcommand{\HCRB}{\ensuremath{\mathrm{HCRB}}}
\newcommand{\bchi}{\boldsymbol{\chi}}
\newcommand{\Rvec}{\mathbf{R}}
\newcommand{\ii}{\mathrm{i}}
\newcommand{\dd}{\mathrm{d}}
\newcommand{\T}{\mathsf{T}}
\newcommand{\diag}{\operatorname{diag}}
\newcommand{\Var}{\operatorname{Var}}
\newcommand{\tr}{\operatorname{tr}}
\newcommand{\defeq}{\vcentcolon=}

\begin{document}
\title{Gaussian Quantum Metrology with Realistic Linear Sensors}

\author{Jacques Ding}
\email{ding@apc.in2p3.fr}
\affiliation{Laboratoire Astroparticule et Cosmologie, Universit\'e Paris Cit\'e, CNRS, F-75013 Paris, France}

\author{James W. Gardner}
\affiliation{Pritzker School of Molecular Engineering, University of Chicago, Illinois 60637, USA}

\author{Tuvia Gefen}
\affiliation{Racah Institute of Physics, The Hebrew University of Jerusalem, Jerusalem 9190401, Israel}

\author{Yanbei Chen}
\affiliation{Theoretical Astrophysics, Walter Burke Institute for Theoretical Physics, California Institute of Technology, Pasadena, California 91125, USA}

\date{\today}

\begin{abstract}
    Quantum sensing promises enhanced precision, but the usual quantum Cramer Rao bound can be too optimistic for realistic linear sensors, where squeezing, filtering, and loss reshape quantum noise. We derive the tight Holevo Cramer Rao bound and show that realistic degradation yields a hierarchy with the usual bound and homodyne readout. This hierarchy already exists in gravitational-wave detectors. We propose a hardware-efficient readout that reaches the Holevo bound without extra signal loss, increasing compact-binary detection rates by up to $25\%$ over the present LIGO homodyne readout.
\end{abstract}

\maketitle

\paragraph{Introduction ---} Linear quantum sensors encode classical input waveforms in continuous-variable quantum output fields. Their sensitivity is set not only by the total quantum noise, but also
by how this noise is distributed among conjugate quadratures and sidebands.
Gravitational-wave (GW) interferometers are a prominent example: each Fourier
component of the strain waveform has two real sine/cosine amplitudes, so the measurement is
intrinsically multiparameter.

The commonly quoted Quantum Cramer Rao bound (\QCRB) optimizes the readout for
each parameter separately. It is therefore a lower bound, but it is
physically attainable only when its optimal estimators are compatible. The relevant attainable bound in a multiparameter problem is the
Holevo Cramer Rao bound (\HCRB), which optimizes over joint measurements. In
ideal tuned, lossless models often used for gravitational-wave interferometers,
this distinction is invisible: the \QCRB, the \HCRB, and variational readout
-- the frequency-dependent homodyne strategy used to evade radiation-pressure
backaction -- all coincide
\cite{Helstrom1969,Helstrom1976,KLMTV,Buonanno_2001,Buonanno_2002,DanKhal12,
    Miao_2017_towards,Gardner_2024,Gefen2024}. However, the LIGO and Virgo detectors,
starting with the fourth observing run (O4), include frequency-dependent
squeezing \cite{aligo_FDS,Virgo_FDS}, which comes with optical loss, mismatch,
and dephasing \cite{Kwee2014,McCuller21}.
These effects can unbalance the quantum noise distribution over the positive and negative frequency sidebands, so it is no longer evident that the usual
QCRB/variational-readout picture gives the attainable quantum limit.

In this Letter, we derive the exact spectral \HCRB\ for tuned Gaussian linear
sensors with realistic degradation. We find a hierarchy controlled by the
output sideband imbalance: for weak imbalance, the \QCRB\ is already
unattainable while variational readout still reaches the \HCRB; for strong
imbalance, all three bounds separate and the optimum becomes nonprojective (see \Cref{tab:hierarchy}). We
identify the attaining measurements and introduce symplectodyne readout, a
narrowband implementation based on local-oscillator modulation rather than
additional lossy optics in the signal path. Applied to the true, full LIGO O4 quantum
noise budget \cite{aligo_FDS,LIGO_SQL,Capote_2025}, the hierarchy is already
present and leads to measurable astrophysical gains.

\begin{table}[t]
    \centering
    \scriptsize
    \setlength{\tabcolsep}{2.5pt}
    \renewcommand{\arraystretch}{1.08}
    \begin{tabular*}{\columnwidth}{@{\extracolsep{\fill}}cccc@{}}
        \toprule
        \textbf{Loss} & \textbf{Imbalance} & \textbf{Hierarchy} & \textbf{Attainability} \\
        \midrule
        No  & $\Delta = 0$    & $\mathrm{QCRB} = \mathrm{HCRB} = \mathrm{Var}$ & PVM \\
        Yes & $\Delta = 0$               & $\mathrm{QCRB} = \mathrm{HCRB} = \mathrm{Var}$ & PVM \\
        Yes & $0 < |\Delta| \le \tfrac{1}{2}$ & $\mathrm{QCRB} < \mathrm{HCRB} = \mathrm{Var}$ & PVM \\
        Yes & $|\Delta| > \tfrac{1}{2}$   & $\mathrm{QCRB} < \mathrm{HCRB} < \mathrm{Var}$ & POVM \\
        \bottomrule
    \end{tabular*}
    \caption{Hierarchy between the QCRB, HCRB, and variational readout (Var) for tuned detectors, depending on the presence of loss, whether they create sideband imbalance, and which regime -- projective (PVM) or non-projective (POVM) -- the optimal measurement belongs to.}
    \label{tab:hierarchy}
\end{table}

\paragraph{Multiparameter estimation in linear sensors ---}
Consider a classical signal (e.g. a GW strain component) $h(t)$, coupled to a sensor (e.g. a GW interferometer). In the linear regime, $h$ can be decomposed along its Fourier components $h[\Omega]$, and it is thus sufficient to focus on the estimation problem at a given sideband frequency \(\Omega\). Under the standard input-output formalism for linear time-invariant systems \cite{GardZoll99,DanKhal12,Miao_2017_PRA,ding2024},
the two-photon output quadratures at frequency \(\Omega\), written as \(x^{\rm out}[\Omega]=(q^{\rm out}[\Omega],p^{\rm out}[\Omega])^\T\), are given by
\begin{equation}
    \begin{split}
        { x}^{\rm out}[\Omega]
        &= H_{\rm }[\Omega]\,{ x}^{\rm in}[\Omega]
        + \sum_j T_j[\Omega]\,{ n}_j[\Omega] + \bchi[\Omega] h[\Omega].
        %&=\delta x^{\rm out}[\Omega]+\bchi[\Omega]\,h[\Omega]
    \end{split}
    \label{eq:bout}
\end{equation}
Here \(\bchi[\Omega]\) is the signal response, while \(x^\text{in}\) and
\(n_j\) are the quantum input and loss-port fields respectively. The transfer matrices \(H\)
and \(T_j\) depend on the loss mechanisms \cite{Kwee2014,McCuller21}, but are constrained so that
\(x^\text{out}\) obeys the canonical commutation relations \cite{ding2024}. For zero-mean input and loss fields, all quantum fluctuations
can be grouped into an output-referred noise \(\delta x^\text{out}\), giving
\begin{equation}
    \begin{aligned}
        x^{\rm out}[\Omega] & =\delta x^{\rm out}[\Omega]+\bchi[\Omega]\,h[\Omega]
    \end{aligned}
    \label{eq:main-gw-io}
\end{equation}

We introduce the normalized cosine/sine quadratures \(\theta_1\) and \(\theta_2\) of the waveform, such that $h(t) = \frac{2}{\sqrt{T} ||\boldsymbol{\chi}||}[ \theta_1 \cos(\Omega t) + \theta_2 \sin(\Omega t)]$ where $T$ is the measurement time. Estimating $h(t)$ is therefore equivalent to simultaneously estimating \(\theta=(\theta_1,\theta_2)^\T \in \mathbb{R}^2\). At each \(\Omega\), the signal therefore labels a displaced Gaussian state
\(\rho_\theta\). A readout is a POVM \(M(\dd y)\) such that
\[
    p_\theta(\dd y)=\tr[\rho_\theta M(\dd y)],
    \qquad
    \int \tilde\theta_i(y)\,p_\theta(\dd y)=\theta_i .
\]
Thus \(\tilde\theta_i\) are unbiased estimators obtained from a quantum
measurement. We quantify their performance by
the usual two-sided quantum noise strain power spectral density (PSD) of the estimation error
$\delta \tilde h = \tilde h - h$ \cite{Maggiore2008,KLMTV}:
\begin{equation}
    \Sh[\Omega]\defeq  \frac{\langle \{ \delta \tilde h[\Omega], \delta \tilde h[\Omega]^\dagger \}\rangle}{2T} =\frac{\Var(\tilde\theta_1)+\Var(\tilde\theta_2)}{\|\bchi[\Omega]\|^2}
    \label{eq:main-Sh}
\end{equation}

For the Gaussian output states considered here, their quantum noise is fully characterized by the hermitian positive definite, spectral density matrix $\bar S[\Omega]$ of the output quadratures \cite{DanKhal12,Miao_2017_towards} which we decompose as \cite{ding2024}
\begin{equation}
    \begin{split}
        \bar S[\Omega]&\defeq \frac{\langle \{ \delta x^\text{out}[\Omega], \delta x^\text{out}[\Omega]^\dagger \}\rangle}{2T}\\
        &=R(\phi)\,S(r)
        \begin{pmatrix}
            \Sigma   & i\Delta \\
            -i\Delta & \Sigma
        \end{pmatrix}
        S(r)\,R(\phi)^\T.
    \end{split}
    \label{eq:main-sbar}
\end{equation}
Here \(R(\phi)\) is a rotation matrix, \(S(r)=\diag(e^r,e^{-r})\), and the real
frequency-dependent parameters obey
\(\Sigma\ge1/2\) and \(\Sigma-|\Delta|\ge1/2\). The parameters \(r,\phi\)
describe output-referred squeezing, while \(\Sigma\pm\Delta\) are the effective
thermal variances of the \(\pm\Omega\) sideband modes after removing that
squeezing. Thus \(\Delta\neq0\) is a sideband-noise imbalance, which drives the
separation between the QCRB, HCRB, and variational readout below.

\begin{figure}
    \centering
    \includegraphics[width=\linewidth]{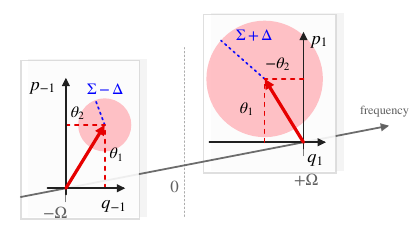}
    \includegraphics[width=0.98\linewidth]{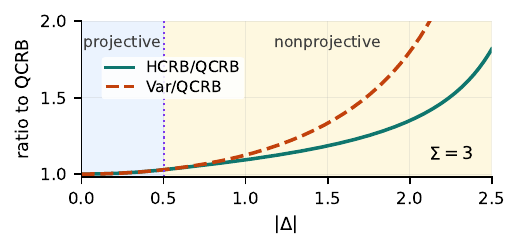}

    \caption{\textit{Top}: Phase space representations of the output quantum state in the sideband modes (see End Matter). For clarity, we set here $r =  \phi = 0$. \textit{Bottom}: Ratios \(\Sh^{\HCRB}/\Sh^{\QCRB}\) and \(\Sh^{\mathrm{var}}/\Sh^{\QCRB}\) as functions of \(\lvert\Delta\rvert\) for $\Sigma = 3$. In the projective branch \(0<\lvert\Delta\rvert\le 1/2\), one has $\Var = \HCRB > \QCRB$, while in the nonprojective branch \(\lvert\Delta\rvert>1/2\), $\Var > \HCRB >\QCRB$.}
    \label{fig:phase-space_representations}
\end{figure}

In the rest of this paper, we focus on tuned linear sensors, where the normalized response is \(\boldsymbol{\chi}[\Omega]\propto (0, 1)^\T\), so the signals sit entirely in the phase quadrature in the two-photon picture. Equivalently, in the single-photon picture, $\theta_1$ ($\theta_2$) sits (a)symmetrically in quadratures of the $\pm\Omega$ sidebands (see \Cref{fig:phase-space_representations} and End Matter).

\paragraph{Variational readout ---}We first review homodyne detection and variational readout \cite{Vyatchanin_Zubova_1995,KLMTV,DanKhal12}.
Homodyne detection measures an output quadrature
\(u^\T x^\text{out}[\Omega]\) with
\(u=(\cos\zeta,\sin\zeta)^\T\), $\zeta \in [0, 2\pi)$.
Variational readout optimizes \(\zeta\) at each frequency, giving
(\Cref{eq:main-Sh,eq:main-sbar})%\james{Eqs.~\ref{eq:main-Sh},~\ref{eq:main-sbar} used?}
\begin{equation}
    \Sh^{\rm var}[\Omega] = \min_{\zeta \in [0,2\pi)}\frac{u^\T \bar S[\Omega] u}{|u^\T \bchi[\Omega]|^2}.
    \label{eq:variational-readout-spectrum}
\end{equation}
This minimization can be carried out exactly \cite{SupplementalMaterial} and yields the closed-form expression
\begin{equation}
    \Sh^{\rm var}[\Omega] = \frac{\Sigma}{g\|\bchi\|^2}
    \label{eq:main-varreadout-compact}
\end{equation}
where all parameters on the right hand side are evaluated at $\Omega$ and \(g\defeq e^{2r}\cos^2\phi+e^{-2r}\sin^2\phi\) is the signal gain due to the output-referred squeezing.

\paragraph{Quantum Cramer Rao Bound ---}
The \QCRB\ \cite{Miao_2017_PRA,Miao_2017_towards} is the single-parameter quantum limit obtained by optimizing the readout separately for each component \(\theta_i\).
The calculation \cite{SupplementalMaterial} yields
\begin{equation}
    \label{eq:main-qcrb-compact}
    \Sh^{\QCRB}[\Omega] := \frac{\Sigma^2-\Delta^2}{g\Sigma\|\bchi\|^2}.
\end{equation}
In a multiparameter problem, however, this lower bound need not be attainable, because the individual optimal estimators $\tilde \theta_1^\text{QCRB}, \tilde \theta_2^\text{QCRB}$ (see End Matter) can be incompatible. Indeed, for lossy systems, the corresponding symmetric logarithmic derivatives do not commute in general \cite{Demkowicz_Dobrza_ski_2020,Holevo2011}, even though the parameter generators do \cite{imai2026hierarchysaturationconditionsmultiparameter}. Here, they suffer from an incompatibility
\begin{equation}
    [\tilde \theta_1^\QCRB, \tilde \theta_2^\QCRB]
    =
    -\,\ii\,\frac{\Delta}{g\Sigma},
    \label{eq:main-qcrb-incompatibility}
\end{equation}
which does not vanish in the general $\Delta \neq 0$ case.

\paragraph{Holevo Cramer Rao Bound ---}
The \HCRB\ resolves this incompatibility issue: it minimizes the error over all jointly measurable estimators \([\tilde \theta_1, \tilde \theta_2] = 0\) and is therefore the attainable multiparameter quantum limit \cite{Holevo2011}.
It takes the compact form \cite{SupplementalMaterial}
\begin{equation}
    \Sh^{\HCRB}[\Omega] := \frac{1}{g\|\bchi\|^2}\left[\Sigma-\frac{(\text{ReLU}(2\lvert\Delta\rvert-1))^2}{4\Sigma}\right]
    \label{eq:main-hcrb-compact}
\end{equation}
where the rectified linear unit \cite{bach2024learning} is defined as \(\text{ReLU}(x)\defeq \max(x,0)\). We call the optimum projective when it is attainable by a projective measurement (PVM) on the signal sideband modes, possibly after active Gaussian processing. We call it \emph{nonprojective} when no PVM on the signal modes attains it, so an ancilla-assisted POVM is required, such as heterodyne detection after active Gaussian processing. As we shall see below, the ReLU term is precisely the improvement over projective strategies which is available only in that latter case: it vanishes when \(\lvert\Delta\rvert\le 1/2\) and activates when \(\lvert\Delta\rvert>1/2\) to lower the HCRB (see \Cref{tab:hierarchy}).

Comparing \Cref{eq:main-hcrb-compact,eq:main-qcrb-compact}, one has
\(\Sh^{\QCRB}\le\Sh^{\HCRB}\), with equality only for \(\Delta=0\).
Physically, \(\Delta\neq0\) means that the two sidebands are unequally noisy.
The separate QCRB estimators weight the cleaner
sideband more strongly. Since that same sideband carries \(\theta_1\) and
\(\theta_2\) in conjugate quadratures, these separate optima are not jointly
measurable.

The HCRB is then the best joint compromise. For
\(0<|\Delta|\le1/2\), the optimum remains projective and coincides with
variational readout, \(\Sh^{\HCRB}=\Sh^{\rm var}\). For
\(|\Delta|>1/2\), an ancilla-assisted POVM becomes advantageous and
\(\Sh^{\QCRB}<\Sh^{\HCRB}<\Sh^{\rm var}\). In the extreme imbalance limit
\(\Sigma\to\infty\), \(|\Delta|\to\Sigma-1/2\), the HCRB tends to the universal bound of twice the QCRB \cite{Carollo_2019_quantumness,Tsang_2020_semiparametric}. \Cref{fig:phase-space_representations} illustrates this hierarchy for
\(\Sigma=3\).

\paragraph{Attainability of the HCRB ---}
For Gaussian shift models, the HCRB (\Cref{eq:main-hcrb-compact}) is not only a bound: it is exactly attainable by a suitable Gaussian measurement on the single-copy level, so it gives the tight spectral limit {\cite{Holevo2011,Demkowicz_Dobrza_ski_2020}}.
We now identify the ideal broadband measurement that attains it, and then a narrowband implementation requiring only a modulated local oscillator.

\begin{figure}
    \centering
    \includegraphics[width=\linewidth]{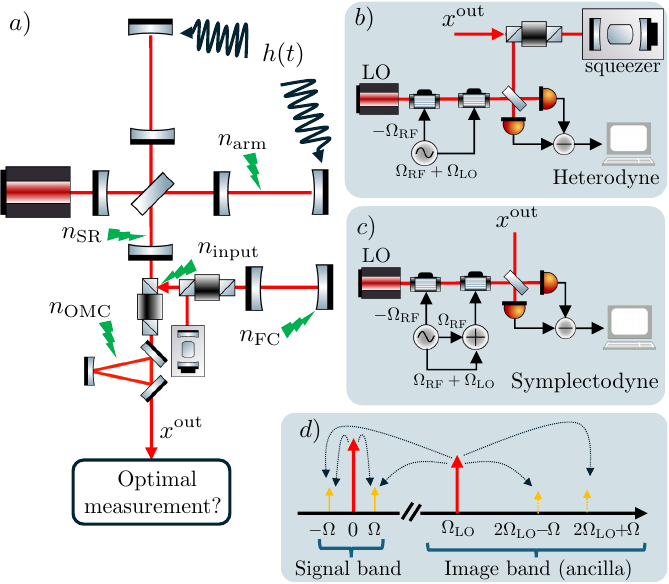}
    \caption{(a) Simplified scheme of LIGO with frequency-dependent squeezing and examples of realistic losses (Filter cavity $n_{\rm FC}$, dark port input $n_{\rm input}$, signal recycling cavity $n_{\rm SR}$, arm cavities $n_{\rm arm}$, output mode cleaner $n_{\rm OMC}$). (b)  HCRB-optimal POVM: the output undergoes Gaussian processing (squeezer), then is detected using heterodyne at frequency $\Omega_\text{LO}$ outside of the signal band ($\sim$ MHz for LIGO). (c,d) Symplectodyne readout: a three-channel RF generator, combined with a pair of upshift/downshift Acousto-Optic Modulators, creates a two-tone local oscillator (carrier frequency and sideband $\Omega_\text{LO}$); physical Gaussian processing is avoided and image-band ancilla modes are automatically included.}
    \label{fig:readout-schemes}
\end{figure}

In the projective regime (\(\lvert\Delta\rvert\le 1/2\)), we have already seen that variational readout followed by sine/cosine demodulation to construct $\tilde h[\Omega]$ is optimal. In the nonprojective regime (\(\lvert\Delta\rvert>1/2\)), the optimal measurement corresponds to a \textit{variational Gaussian processing} built from frequency-dependent squeezers, followed by heterodyne detection \cite{SupplementalMaterial} (see \Cref{fig:readout-schemes}(b) for illustration using a single squeezer).
Because \(\Delta[\Omega]\) can cross the projective/nonprojective threshold as a function of frequency, an ideal broadband implementation would first spatially separate the two regimes, for example with a Fabry--Perot spectral demultiplexer.

A simpler narrowband route is \emph{symplectodyne} readout: balanced detection with a two-tone local oscillator (LO), followed by classical post-processing of the photocurrent harmonics (see \Cref{fig:readout-schemes}(c-d)). This synthesizes the same effective squeezed-heterodyne statistic without requiring a physical squeezer. The signal-free image sidebands supply the vacuum ancilla required by the optimal POVM. Adding more tones in the LO gives no broadband advantage \cite{SupplementalMaterial}.

\paragraph{Application to LIGO O4 ---}
We now apply the theory to the measured LIGO quantum-noise budget in O4, taken from \cite{LIGO_SQL}. In
\Cref{eq:bout}, \(x^\text{in}\) is the injected squeezed vacuum, and
the \(n_j\) are vacuum modes describing optical losses at different points of the propagation through the interferometer \cite{Kwee2014} (see \Cref{fig:readout-schemes} (a) for a visual representation of the interferometer and some of the loss terms).  We obtain \(H,T_j,\bchi\) (which include mode mismatch effects) at each frequency using the
\texttt{gwinc} Python package \cite{pygwinc_ascl}, then extract \(r,\phi,\Sigma,\Delta\)
from \Cref{eq:main-sbar}. Phase noise is incorporated in these parameters as described in \cite{SupplementalMaterial}. Although the signal extraction cavity is slightly detuned \cite{Capote_2025,LIGO_SQL}, the response is
well approximated by the tuned condition \(\bchi\propto(0,1)^\T\) so that our general closed-form expressions can be applied \cite{SupplementalMaterial}. In all comparisons below, the interferometer and squeezing parameters are fixed to the O4 configuration; only the readout is varied for a fair comparison. 

\begin{figure}[tbp]
    \centering
    \includegraphics[width=\linewidth]{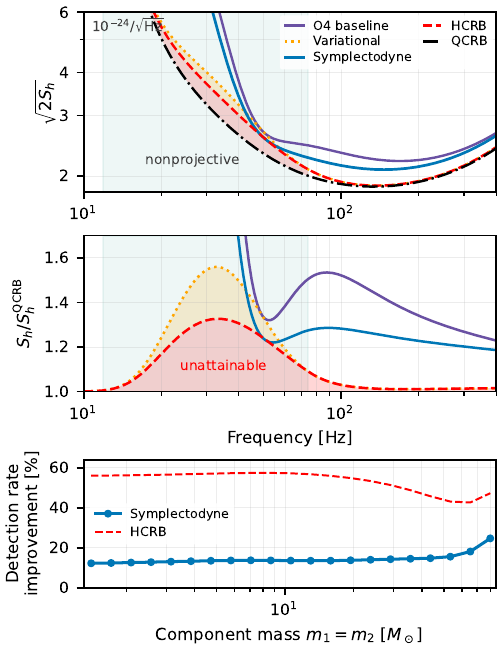}
    \caption{LIGO O4 quantum-noise hierarchy and astrophysical impact. \textit{Top}: One-sided quantum strain amplitude spectral density, for baseline LIGO O4, variational readout, symplectodyne optimized at \(\SI{50}{Hz}\), HCRB, and QCRB. \textit{Middle}: PSD ratios to the QCRB; yellow fills mark the region where symplectodyne has an advantage over homodyne; red fill marks the unattainable QCRB--HCRB gap. \textit{Bottom}: Relative detection-rate improvement over baseline O4 for equal-mass binaries, using the full O4 classical plus quantum noise budget, with the optimal symplectodyne readout or the HCRB noise floor.}
    \label{fig:o4-readout-family}
    \label{fig:o4-detection-rate}
\end{figure}

The extracted imbalance crosses the nonprojective threshold,
\(|\Delta(\Omega)|>1/2\), from \SI{12}{Hz} to \SI{74}{Hz}
(\Cref{fig:o4-readout-family}). In this band the Homodyne-HCRB-QCRB hierarchy is strict, with peak PSD ratios
\(\max(\Sh^{\HCRB}/\Sh^{\QCRB})\approx 1.3\) and
\(\max(\Sh^{\mathrm{var}}/\Sh^{\QCRB})\approx 1.6\). The imbalance is driven
by the combination of high generated squeezing (\(17 \, \mathrm{dB}\), compared to \(6\,\mathrm{dB}\) measured at the output) and filter-cavity detuning mismatch, which
rotates and dephases the squeezed field near the
\(\SI{26}{Hz}\) filter-cavity detuning \cite{Kwee2014,McCuller21,ding2024,Korobko_2026,Kuns_Brown_2026}.

We also show, in the top and middle plots of \Cref{fig:o4-readout-family}, the broadband performance of a symplectodyne readout optimized to
reach the HCRB at \(\SI{50}{Hz}\), which outperforms variational readout in the
\(43-\SI{57}{Hz}\) band, and outperforms the homodyne-type O4 readout \cite{Hild_2009,Fricke_2012,Capote_2025} over the whole detection band. Its practical advantage is that the required
operation is transferred to the local oscillator: no additional optical element
is inserted in the signal path, so the output field is not exposed to extra
readout loss. By contrast, broadband variational readout and variational Gaussian processing require low-loss
kilometer-scale output filter cavities and squeezers in the signal path, making its
performance highly sensitive to optical loss \cite{KLMTV}.

As a simple astrophysical figure of merit, we add the standard O4 classical noise budget (seismic, thermal, residual gas, etc.) to the quantum noise spectrum and convert into inspiral range and hence detection rate for equal-mass compact binaries. \Cref{fig:o4-detection-rate} (bottom plot) summarizes the improvements in detection rate relative to the baseline LIGO O4 configuration for equal mass compact binary mergers. With the full O4 classical and quantum noise budget, optimized symplectodynes for each component mass provide a \(12\text{--}25\%\) improvement compared to the homodyne-type readout experimentally implemented in LIGO O4, while the HCRB sets a \(43\text{--}57\%\) ceiling for the same baseline. An adaptive symplectodyne readout could reach this ceiling in the adiabatic limit, by tuning the LO modulation across the instantaneous frequency trace of the chirped signal.

The symplectodyne gain is largest for the heaviest binaries, which weight the nonprojective band most strongly. This mass range is astrophysically motivated: binaries with component masses \(\gtrsim 50\,M_\odot\) probe the onset of the pair-instability regime and likely hierarchical mergers, while events such as GW190521 already show that they can produce intermediate-mass-black-hole remnants \cite{GW190521,Tong_2026,Torniamenti_2024}. Thus improved low-frequency quantum readouts are especially relevant for formation-channel studies and black-hole spectroscopy \cite{Torniamenti_2024,Carullo_2025}.

\paragraph{Conclusion ---}

In this Letter, we developed a general framework for Gaussian quantum waveform estimation in
tuned linear sensors with realistic degradation. The result is a compact
closed-form hierarchy for variational readout, the QCRB, and the attainable
HCRB, that arises only when sideband
imbalance is properly considered. This imbalance, missed by the previous simplified models \cite{Miao_2017_towards,Gardner_2024,Miao_2019}, is what
separates the ideal single-parameter QCRB from the true multiparameter quantum limit even when the sensor is tuned.

Applying the framework to the LIGO O4 quantum-noise budget, we found that the
detector already operates in a nonprojective band where
\(\Sh^{\QCRB}<\Sh^{\HCRB}<\Sh^{\mathrm{var}}\). The optimal readout is therefore
not ordinary homodyne detection, but an ancilla-assisted Gaussian measurement.
Our symplectodyne construction realizes this measurement narrowband by shaping
the local oscillator rather than inserting lossy optics in the signal path. 
Additionally, unlike synodyne detection \cite{Buchmann_2016,Ockeloen_Korppi_2018}, which causes
frequency folding and downconversion to noise-prone low frequencies,
symplectodyne places its auxiliary tones outside the GW band. This makes it a
natural candidate for future upgrades and next-generation detectors \cite{Capote_2025,Srivastava_2022,Branchesi_2023}, where larger generated
squeezing and lower classical noise should increase the advantage of mitigating dephasing noise and accessing the HCRB.

In our analysis, we fixed the input squeezed state and optimized the measurement. A complementary state-preparation approach is to adjust the input squeezing level to decrease dephasing noise; this can also lead to an enhancement in detection rate \cite{SupplementalMaterial}, but changes the quantum statistical model and therefore the relevant quantum Cramer--Rao and Holevo bounds. The advantage of symplectodyne detection compared to squeezing input optimization is that the local oscillator is RF-modulated electronically, while the bandwidths of the filter cavity and the interferometer are limited by the kHz-band signal frequencies. Thus symplectodyne can in principle be controlled much faster, which may enable adaptive readout after early warning of gravitational-wave signals.

Beyond gravitational-wave interferometry, our derivation gives a rare analytic
example of an attainable multiparameter quantum bound for a mixed-state
Gaussian model. General HCRB calculations usually require optimizing over
density-operator geometry or solving semidefinite programs \cite{Holevo2011,Albarelli_2019_Holevo,Sidhu_2021}, which may obscure the intuition of the underlying physical mechanisms. Here the conjugate-symplectic reduction of \Cref{eq:main-sbar} turns the problem
into elementary matrix algebra, making the role of sideband imbalance and
measurement incompatibility explicit. In this Gaussian-shift setting, we note that the HCRB
also coincides with the most-informative bound \cite{Conlon_2025,yung2025}.

Future work should quantify technical noise from LO modulation, and explore
whether non-Gaussian input states, which can reduce incompatibility in joint
quadrature-displacement estimation \cite{Frigerio_2025_nongaussian,labarca2026quantum}, can further
improve the attainable waveform limit. Astrophysical figures of merit other than the inspiral range, such as parameter-estimation accuracy, early-warning time, or population-dependent rates, may also lead to different optimal quantum schemes.

\paragraph{Acknowledgments}
We thank Lee McCuller for useful comments regarding the manuscript.
We thank Wenxuan Jia \& Kevin Kuns for insightful discussions regarding the GWINC Python package and the LIGO O4 noise budget. We acknowledge helpful contributions from OpenAI GPT 5.2 \& 5.4 on the derivations of the HCRB (\Cref{eq:main-hcrb-compact}) and the optimality of the symplectodyne readout. The authors are fully responsible for the scientific content and integrity of this work.
J.W.G.\ acknowledges support from the ARO (W911NF-23-1-0077), ARO MURI (W911NF-21-1-0325), AFOSR MURI (FA9550-21-1-0209, FA9550-23-1-0338), DARPA (HR0011-24-9-0359, HR0011-24-9-0361), NSF (ERC-1941583, OMA-2137642, OSI-2326767, CCF-2312755, OSI-2426975), Packard Foundation (2020-71479), and DOE (DE-AC02-06CH11357).
T.G. acknowledges funding from ISF Grant No. 3302/25, and the quantum science and technology early-career grant of the Israeli council for higher education.
Y.C. acknowledges support from the Simons Foundation (Award No. 568762) and the National Science Foundation (via Grants No. PHY-2309211 and No. PHY-2309231).

\bibliography{refs}

\begin{widetext}
    \section{End Matter}
    \subsection{Two-photon and sideband conventions}
    The physical sideband (or single-photon picture) modes at $\pm \Omega$ have annihilation operators $a[\pm \Omega]$ satisfying $[a[\Omega], a[\Omega']^\dagger]=2\pi \delta[\Omega -\Omega']$. For a finite measurement time $T \simeq 2\pi \delta[0]$, we regularize these continuous-frequency modes by the bins $a_{\pm 1} = {a[\pm \Omega]}/{\sqrt{T}}$, thus satisfying $[a_m,a_n^\dagger]=\delta_{mn}$ $(m,n=\pm1)$. The quadrature vector $\Rvec=(q_{+1},p_{+1},q_{-1},p_{-1})^\T$ is such that
    \begin{equation}
        q_{\pm 1} = \frac{a_{\pm 1} + a_{\pm 1}^\dagger}{\sqrt 2}, \qquad p_{\pm 1} = \frac{a_{\pm 1} - a_{\pm 1}^\dagger}{\ii \sqrt 2}.
    \end{equation}

    In the frequency-domain input-output relation of \Cref{eq:bout}, we introduce the two-photon quadratures \(x[\Omega]=(q[\Omega],p[\Omega])^\T\). They are defined as
    \begin{equation}
        q[\Omega] = \frac{a[\Omega] + a^\dagger[-\Omega]}{\sqrt 2}, \qquad p[\Omega] = \frac{a[\Omega] - a^\dagger[-\Omega]}{\ii \sqrt 2}.
    \end{equation}
    They are not Hermitian, but obey $q[\Omega] = q[-\Omega]^\dagger$ and $p[\Omega] = p[-\Omega]^\dagger$.
    One has the exact conversion
    \begin{equation}
        \Rvec= \frac{1}{\sqrt{T}}
        \begin{pmatrix}
            1 & 0 & 0  & -1 \\
            0 & 1 & 1  & 0  \\
            1 & 0 & 0  & 1  \\
            0 & 1 & -1 & 0
        \end{pmatrix}
        \begin{pmatrix}
            \operatorname{Re}q[\Omega] \\
            \operatorname{Re}p[\Omega] \\
            \operatorname{Im}q[\Omega] \\
            \operatorname{Im}p[\Omega]
        \end{pmatrix}.
        \label{eq:main-sideband-conversion}
    \end{equation}
    Thus the conjugate-symplectic Williamson form of \(\bar S\) in
    \Cref{eq:main-sbar} is equivalently the ordinary \(4\times4\) real Williamson
    form
    \begin{equation}
        V_R=S_R(r,\phi)
        \diag(\Sigma+\Delta,\Sigma+\Delta,\Sigma-\Delta,\Sigma-\Delta)
        S_R(r,\phi)^\T,
        \label{eq:main-real-williamson}
    \end{equation}
    where \(V_R=\langle\{\delta\Rvec,\delta\Rvec^\T\}\rangle/2\) and
    \begin{equation}
        S_R(r,\phi)=
        \begin{pmatrix}
            \cosh r            & 0                   & \sinh r\cos(2\phi) & \sinh r\sin(2\phi)  \\
            0                  & \cosh r             & \sinh r\sin(2\phi) & -\sinh r\cos(2\phi) \\
            \sinh r\cos(2\phi) & \sinh r\sin(2\phi)  & \cosh r            & 0                   \\
            \sinh r\sin(2\phi) & -\sinh r\cos(2\phi) & 0                  & \cosh r
        \end{pmatrix}
        \label{eq:main-real-symplectic}
    \end{equation}
    is a rotated two-mode squeezing matrix.

    When \(\bchi/\|\bchi\|=(0,1)^\T\), as assumed in this Letter, the
    mean displacement on the sideband modes is
    \begin{equation}
        \langle\Rvec\rangle=
        (-\theta_2,\theta_1,\theta_2,\theta_1)^\T .
        \label{eq:main-sideband-encoding}
    \end{equation}
    The top panel of \Cref{fig:phase-space_representations} is the
    \(r=\phi=0\) version of this representation: the \(+\Omega\) and
    \(-\Omega\) sideband modes have thermal variances \(\Sigma+\Delta\) and
    \(\Sigma-\Delta\), respectively, while \(\theta_1\) displaces both phase
    quadratures and \(\theta_2\) displaces the two amplitude quadratures with
    opposite signs.

    \subsection{Optimal estimators}
    The optimal estimators for variational readout, QCRB, and HCRB are given by
    \begin{equation}
        \tilde\theta_1^\bullet+\ii\tilde\theta_2^\bullet
        =
        u_\bullet^\dagger x^\text{out}[\Omega]+\nu_\bullet,
        \qquad
        \bullet\in\{\mathrm{var},\QCRB,\HCRB\},
        \label{eq:main-endmatter-estimator-form}
    \end{equation}
    where
    \begin{align}
        u_{\mathrm{var}}^\dagger
                           & =
        \frac{1}{\sqrt T}
        \left(
        \frac{(e^{-2r}-e^{2r})\sin(2\phi)}{2g},
        1
        \right),
                           &
        \nu_{\mathrm{var}} & =0,
        \label{eq:main-endmatter-u-var}  \\
        u_{\QCRB}^\dagger
                           & =
        \frac{1}{\sqrt T}
        \left(
        \frac{(e^{-2r}-e^{2r})\sin(2\phi)}{2g}
        +\ii\frac{\Delta}{g\Sigma},
        1
        \right),
                           &
        \nu_{\QCRB}        & =0,
        \label{eq:main-endmatter-u-qcrb} \\
        u_{\HCRB}^\dagger
                           & =
        \frac{1}{\sqrt T}
        \left(
        \frac{(e^{-2r}-e^{2r})\sin(2\phi)}{2g}
        +\ii\frac{\operatorname{sgn}(\Delta)\text{ReLU}(2\lvert\Delta\rvert-1)}{2g\Sigma},
        1
        \right),
                           & \nu_{\HCRB}
                           & =
        \sqrt{\frac{\text{ReLU}(2\lvert\Delta\rvert-1)}{2g\Sigma}}
        \left[-\operatorname{sgn}(\Delta)p_a+\ii q_a\right].
        \label{eq:main-endmatter-nu-hcrb}
    \end{align}
    Here \(a_a=(q_a+\ii p_a)/\sqrt2\) is an independent vacuum ancilla mode, with
    \([q_a,p_a]=\ii\) and \(\Var(q_a)=\Var(p_a)=1/2\).
\end{widetext}

\end{document}